# Excitation dependent Raman studies of self-seeded grown InN nanoparticles with different carrier concentration

Kishore K. Madapu,[1,*] S. R. Polaki,[2] Sandip Dhara[1,*]

[1] Nanomaterials and Sensors Section, [2] Nanomaterials and Characterisation Section, Surface and Nanoscience Division, Indira Gandhi Centre for Atomic Research, Kalpakkam–603 102, India.

[*] Electronic Mail: madupu@igcar.gov.in; dhara@igcar.gov.in

**Abstract**

High quality InN nanoparticles are grown using atmospheric chemical vapour deposition technique via a self-seeded catalytic approach in the temperature range of 580 - 650 $^{\circ}$C. In this temperature region, nucleation barrier of InN is overcome by seeding the low density In nano-particles prior to the reactive $NH_3$ introduction. Samples having increasing carrier density are grown with the help of increasing growth temperature to understand its role in optical phonon structure. Near-resonance Raman spectra show complete different phonon pictures as compared to those for the off–resonance spectra. A Raman forbidden mode of $B_1$(high), because of the possible breakdown of selection rule in the near-resonance condition, is invoked for the first time. Intensity and frequency of this mode strongly depend on the carrier concentration in the sample. In off–resonance conditions, $A_1$(LO) mode for the sample with high carrier concentration is dominated by Fano interference rather than plasmon-phonon coupling. Variation of intensity of the $B_1$(high) mode is correlated with band filling effect, which is substantiated by the luminescence studies of the InN samples with different carrier concentration.

## Introduction

InN belongs to the family of III-nitrides along with the other well known materials such as GaN, AlN. However, InN is less studied as compared to other III-nitrides because of inherent difficulties in synthesising the high quality material.[1,2] InN is predicted to possess the highest electronic mobility among the III–V nitrides, 4400 cm$^2$V$^{-1}$s$^{-1}$.[3] In addition to this, theoretically it is predicted that the low field mobility of InN can go up to 10,000 cm$^2$V$^{-1}$s$^{-1}$.[4] Moreover, high saturation velocity and lowest electron effective mass makes the InN ideal candidate for high speed electronic devices. Being a degenerate semiconductor, InN added with its inherent low crystalline quality lead it to a high carrier concentration material, which results in the blue shift of band gap energy from the theoretically predicted band gap value because of the Burstein Moss effect. In recent times, crystal quality of InN is enormously improved with controlled growth techniques such as molecular beam epitaxiy and metal organic chemical vapour deposition.[5–10] Since last decade onwards, the band gap of InN is revised to ~ 0.75 eV and the accepted band gap is less than earlier predicted value of ~ 2 eV. A revised band gap value of InN opened up the new applications in infrared (IR) region as emitters and detectors.[1,2,11] However, high quality material is essential to be produced in the atmospheric pressure chemical vapor deposition (APCVD) technique because of its possible large scale synthesis capability. Recently, InN nanowires were reported to be grown above its decomposition temperature with the mediation of O using the APCVD technique.[12]

InN is reported to have the widest phonon band gap and lowest photonic band gap which prompt the applications in hot carrier solar cells.[13,14] Thus, it is inevitable to study the phonon structure of InN with the excitation in the IR as well as in the visible region for developing the InN based devices. However, band gap of InN falls in the IR region and the device performance is strongly influenced by the resonance phenomenon. As a result, one can expect the dramatic changes in the phonon structure of InN when it is excited with light in the IR region, subsequently for device characteristics. Crystal quality and basic phonon structure of InN are well studied using inelastic light scattering by means of Raman spectroscopy. However, vibrational properties of InN studied by Raman spectroscopy are very much controversial as much as they are studied.[15–21] The controversies in Raman spectroscopic studies of InN surround the phonon structure of longitudinal optical ($A_1$(LO)) mode. As a degenerate semiconductor, one can expect the large influence of carrier concentration on the vibrational properties of InN. In addition to this, InN is a polar semiconductor; electric field associated with LO phonons strongly interacts with the carrier concentration through the Fröhlich interaction leading to the observation of plasmon-phonon coupling modes in the Raman spectrum.[22] One of the two coupling modes is usually observed in the range of 440–450 cm$^{-1}$, which is assigned as the low frequency coupling mode (PLP) mode ($L^-$).[23] As the carrier concentration increases, this mode strengthens and approaches to $A_1$(TO) mode. However, theoretically predicted high frequency coupling mode ($L^+$) ~ 885 cm$^{-1}$,[23] is seldom observed in previous studies. Instead of the $L^+$ mode there is a peak near the unscreened $A_1$(LO) mode frequency, observed frequently in the range of 580–600 cm$^{-1}$. This observation created a lot of curiosity among the researches regarding the origin of the peak in that region. Thus, observation of the unscreened $A_1$(LO) mode is studied and discussed by lot of researchers. Among the first, presence of the depletion region near InN surface was made responsible for the observation of the unscreened $A_1$(LO) mode.[16] However, Kasic *et. al.*[17] ruled out the possible role of surface depletion region since the charge depletion region is narrower than light penetration depth ($\leq$ 5 nm). Moreover, peak near the $A_1$(LO) mode is attributed to the low frequency PLP mode ($L^-$). This mode rises above $A_1$(TO) mode and behaves like pure phonon mode as a result of scattering process which involves the higher value of wave vectors. In contrary to this, Inushima *et. al.*[18] observed a peak in the range of 440–450 cm$^{-1}$, which was assigned as $A_1$(TO) mode because of its passiveness towards varying carrier concentration. In addition to it, a peak in the vicinity of the $A_1$(LO) mode was assigned as the unscreened $A_1$(LO) mode. Similar to $A_1$(TO) mode, the unscreened $A_1$(LO) mode also does not change its behaviour with carrier concentration and does always have the asymmetric line shape. Asymmetric nature of $A_1$(LO) mode was explained by the decoupling of plasmon-phonon coupled modes and Fano interference. In another independent study,[21] it was proved that phonon mode in the vicinity of the unscreened $A_1$(LO) mode must be because of the plasmon–phonon coupling mode by analyzing its nature after and before annealing in the temperature range of 80–700K. However, there are limited numbers of reports on Raman scattering studies in a resonance condition where the Fröhlich interaction is expected to be very strong.[24–26] Thus, one can shed more light on the phonon structure of $A_1$(LO) and other forbidden modes in the near-resonance condition because of its strong coupling nature through Fröhlich interaction.

In the present report, we demonstrated the successful growth InN nanoparticles on sapphire substrate with self seeded catalytic approach. Variation of carrier concentration in grown samples is achieved with the manoeuvring of the growth temperature. Subsequently, we studied the vibrational properties of unintentionally doped InN nanoparticles grown on sapphire substrate using APCVD technique in a self-seeded catalytic approach. We explored the forbidden Raman modes with excitation dependent Raman scattering process. In addition to it, effects of intrinsic carrier concentration as well as thermal excited carriers were studied on the forbidden Raman mode in the near-resonance condition. Moreover, role of the band filling effect on the forbidden mode as a consequence of increased carrier concentration in the samples is also analysed by studying the luminescence properties.

**Experimental section**

InN nano-crystals were prepared in the APCVD technique with a horizontal tube furnace.[27,28] Metallic In (99.999%) and reactive NH$_3$ (99.999%) were used as the precursor materials. Growth of InN in the APCVD technique is limited by the growth temperature because of the low thermal stability of InN. Moreover, InN was reported to be converted into In$_2$O$_3$ above 550 $^o$C in the presence of O.[28] Nucleation of InN phase above 600 $^o$C was also very difficult because of the possible reversible reaction and decomposition. Thus, to overcome the nucleation barrier of InN, a novel method of self-seeded catalytic growth process was adopted in the present study.[29] The In precursor and sapphire substrate were kept with the separation of 1 cm in the downward flow direction. The growth temperature was ramped to 500 °C at the rate of 20 °C/min and was kept for 30 minutes. At this temperature, nucleation sites are seeded with the In nanoparticles (supplementary Fig. S1). Subsequently, temperature was increased to 580, 600, 620, and 650 °C for the growth of samples, represented as A, B, C, and D, respectively.

Morphology of grown materials was studied using the field emission scanning electron microscopy (FESEM, AURIGA, Zeiss). Structural analysis of these nanoparticles was carried using high resolution transmission electron microscopy (HRTEM, Libra 200 FE Zeiss). Vibrational studies are carried using the Raman spectroscopy (Renishaw, inVia) with different excitation sources of 785, 514.5 and 488 nm in the back scattering geometry. For the 785 and 514.5 nm laser excitations, a 1800 gr/mm grating as monochromator was used and a 2400 gr/mm grating was used for the 488 nm laser excitation. For all these cases, dispersed scattered light was detected by the thermoelectric cooled charged coupled device (CCD) detector. An objective of 100X with numerical aperture (N.A.) of 0.95 is used in the measurements. IR region of the photoluminescence (PL) spectra for 785 nm laser excitation were collected by the InGaAs detector using a 600 gr/mm grating as monochromator and a 20X objective.

**Results and discussion**

Figures 1(a)-(d) shows the variation of morphology of nanoparticles with growth temperature. Sample grown at 580 °C shows the random morphological nanostructures (Fig. 1(a)). Similarly, sample grown at 600 °C also shows the random nanostructures (Fig. 1(b)). Growth of protruded hexagonal nanostructures (dotted cricle), in plenty, was observed in case of 620 °C grown sample (Fig. 1(c)). However, 650 °C grown sample shows random nanostructures with no special features. Inset of the Figure 1(a) shows the HRTEM image of the InN nanoparticle grown at 580 °C. The measured spacing of lattice fringes is well matched with the *m*-plane of (1−100) of hexagonal InN.

Vibrational properties are studied by Raman spectroscopy for various excitations. For all excitations, measurements were carried in the backscattering geometry represented by Z(X−)Z, where Z and X represent the directions of incident laser and electric field, respectively. Among the group theory predicted vibrational modes, $\Gamma = 2A_1+2E_1+2E_2+2B_1$, the wurtzite InN have six Raman active modes.[15,30] The $B_1$ phonon mode is not active for Raman as well as IR spectroscopy. Figure 2(a) shows the Raman spectra of InN nanostructures grown at different temperatures with the excitation of 514.5 nm laser at room temperature (RT). For all the samples, distinct peaks in the range of 85–87 and 488–493 cm$^{-1}$ are observed which correspond to the $E_2$(low) and $E_2$(high) modes, respectively. Considerable low full width half maxima (FWHM) of the $E_2$(high) phonon mode reveals the high crystalline quality of the grown material. In addition to these modes, two more peaks are found to dominate the spectra with peak positions (*i*) at around 443 cm$^{-1}$ and (*ii*) in the range of 589–593 cm$^{-1}$ which appear to be forbidden $A_1$(TO) and close to the unscreened $A_1$(LO) modes, respectively. The phonon mode near unscreened $A_1$(LO) mode in the spectra shows different behaviour of asymmetric broadening in the high temperature grown samples (C and D) as compared to that for the low temperature grown samples (A and B). This observation suggests that, instead of the plasmon-phonon coupling modes, Fano interference between carriers and LO phonons dominates in the present case leading to the asymmetric nature of $A_1$(LO) mode.[18] Consequently, the peak observed in the range of 585–595 cm$^{-1}$ is assigned as the unscreened $A_1$(LO) mode. Interestingly, the Fano line shapes are observed only in high temperature grown samples (C and D). However, low temperature grown samples show almost symmetric nature of the mode. Asymmetric broadening of the $A_1$(LO) mode in high temperature grown samples, C and D, is attributed to the increase in the intrinsic charge carrier density.[31] Hence, Raman spectroscopic analysis reveals that the low temperature grown samples are having the low carrier concentration as compared to that for the high temperature grown samples. Increase in carrier concentration in high temperature grown samples, C and D, is attributed to the N vacancies ($V_N$). At high growth temperatures InN nanoparticles are prone to have the large number of $V_N$ because of low thermal stability of InN.[28,29] In addition to this, because of its passiveness towards the carrier concentration, the phonon mode observed at 443 cm$^{-1}$ is assigned as the $A_1$(TO) mode instead of low frequency PLP coupling mode ($L^-$).[23] However, $A_1$(TO) mode is forbidden in the present configuration. Appearance of forbidden $A_1$(TO) mode is attributed to the violation of polarization selection rules as a finite size effect in nanostructures. Figure 2(b) shows the Raman spectra collected with the 488 nm laser excitation at RT. This spectra also show the similar features as observed for 514.5 nm excitation in the context of peak positions and asymmetric broadening of $A_1$(LO) mode for the high temperature grown samples.

The RT Raman spectra of InN nanostructures excited with a 785 nm laser show (Fig. 2(c)) completely different phonon picture, as compared to other excitations of 514.5 and 488 nm. A considerable broad peak was observed in the range of 440–600 cm$^{-1}$. Nevertheless, low temperature grown samples still shows the signature of symmetry allowed modes by peaking at their corresponding positions, indicated in the spectra by α for $A_1$(TO), β for $E_2$(high), and γ for $A_1$(LO) modes. An additional broad peak is also observed in the range of the 540–560 cm$^{-1}$, as indicated by δ (Fig. 2(c)). No symmetry allowed Raman mode appears in this frequency region for wurtzite InN. A tiny phonon mode in this region is also reported by Inushima *et al.*[18] without any assignment. On the other hand, the $B_1$(high) mode which is forbidden in wurtzite InN, is reported around 550 cm$^{-1}$.[1] In addition to this, Thakur *et al.*[19] also observed a peak in this region, assigned to be the $B_1$(high) mode with the excitation of 785 nm where peak intensity varies with carrier concentration in the system. Thus the broad peak observed in the range of the 540–560 cm$^{-1}$, in the present study, is assigned as the $B_1$(high) mode. Nevertheless, so far adequate attention is not spent in the assignment of the forbidden $B_1$(high) mode. In the present case, appearance of forbidden mode is attributed to the near-resonance conditions invoking breakdown of the selection rule.[32] However, structural disorder in the material can also induce the forbidden modes in between TO and LO modes.[25] If the forbidden mode is originated because of structural disorder then one can expect the same mode to appear in the Raman spectra excited with higher energies (514.5 and 488 nm) also. However, we failed to observe the $B_1$(high) forbidden mode in the high energy excited Raman spectra (Figs. 2(a),(b)). This observation substantiates the fact that the $B_1$(high) forbidden mode, which appears in the Raman spectrum excited with 785 nm (Fig. 2(c)), is because of the near-resonance conditions and is not activated because of disorder in the material. The peak position, as well as intensity of the $B_1$(high) mode are found to vary with the carrier concentration in the sample. In case of high temperature grown samples (C and D), the peak is red shifted as compared to that for the low temperature grown samples (A and B) (indicated by dashed vertical line in Fig. 2(c)). Moreover, the peak intensity of $B_1$(high) mode is found to be strengthened enormously in 650 °C grown sample (D), dominating over the other Raman modes. Red shift of the $B_1$(high) mode with the increasing growth temperature elucidates the fact that the phonon mode is strongly dependent on the carrier concentration in the material. Effect of temperature on the resonance induced forbidden mode, $B_1$(high) was studied using a long working distance 50X objective with N.A. of 0.45. Variation of sample temperature was carried out using the liquid $N_2$ cooled low temperature stage (Linkam T95, UK). The temperature was varied in the range of 80–373 K. Figure 3(a) shows the Raman spectrum of 580 °C grown sample (A) measured at 80 K with proper background correction. The active region of the spectrum is de-convoluted with Gaussian function. The spectrum shows $B_1$(high) mode at ~ 565 cm$^{-1}$.[1,18,19] The ratio of the intensity of $B_1$(high) and $A_1$(LO) phonon mode [I($B_1$(high) / I($A_1$(LO)] was calculated to be 1.26. In contrary to the Raman spectrum measured at 80K, $B_1$(high) mode was observed to red shift at ~ 544 cm$^{-1}$ for the same sample measured at 373 K (Fig. 3(b)). However, I($B_1$(high) / I($A_1$(LO) was calculated to be significantly higher value of 1.60 when measured at 373K. Similar kind of measurements was also carried out for high temperature grown sample, D. Figures 3(c) and 3d show the Raman spectra measured at 80 K and 373 K, respectively. The frequency of $B_1$(high) mode was found to be red shifted from 552 to 536 cm$^{-1}$ when measurement temperature is increased from 80 to 373K. The intensity ratio [I($B_1$(high)) / I($A_1$(LO)] was found to be 1.76, and 1.98 when measured at 80 and 373K, respectively. In other words, intensity of the forbidden mode is always sufficiently stronger than that for the $A_1$(LO) phonon mode. Interestingly, the intensity ratio [I($B_1$(high)) / I($A_1$(LO)] for the sample D when measured at 80K was found to be stronger than that for the sample A, at all temperatures. This observation suggests that high temperature grown samples have sufficiently higher carrier concentration which makes the resonance induced forbidden mode, $B_1$(high) stronger even at low temperature.

Raman spectra collected at different temperatures from 80–373 K for the sample A is shown in Figure 4(a). Domination of the $B_1$(high) mode over $A_1$(LO) phonon mode was observed to initiate at 200K. Above the flipping temperature of 200K, the $B_1$(high) mode is found to dominate over the $A_1$(LO) phonon mode. This observation elucidates the fact that thermally generated carriers increase with the increasing temperature, leading to the increase in the intensity of the forbidden mode, $B_1$(high). Thus, the intensity of the $B_1$(high) mode is influenced even by the thermally generated carriers. Similar kind of trend is also observed in case of the sample grown at 600 $^o$C (not shown in figure). The temperature dependent measurements on the sample D shows (Fig. 4(b)) that the $B_1$(high) is always dominant over $A_1$(LO) phonon mode at all measurement temperatures.

Luminescence emission studies are carried out to understand the variation of forbidden mode intensity with respect to the carrier concentration of the samples. Figure 5(a) shows the PL spectra with a broad emission in the range of 1.2–1.4 eV for InN samples grown at different temperature with an excitation of 785 nm. Peaks indicated by the (∗) in the spectra correspond to the first (~530 cm$^{-1}$ at ~1.513 eV in Fig. 5(a)) and second order (1189 cm$^{-1}$ at ~1.432 eV in Fig. 5(a)) Raman features of the InN riding on the PL emission. The emission peak is observed to blue shift with increasing growth temperature, in the range of 580–650 $^o$C.

However, this broad emission cannot be attributed to the band edge emission. The PL emission of the samples A and D, collected using the InGaAs detector with the excitation of 785 nm and 600 gr/mm grating, shows band edge peaks in the range of 0.8−1.2 eV (Fig. 5(b)). The shift in the PL spectra (Fig. 5(b)) from the reported band gap value of 0.75 eV in bulk InN, can be understood because of the combined effect of Burstein Moss shift as well strain in the nanoparticles.[29] Thus, the interesting observation of broad and relatively sharp band edge emission using different detectors, can be understood in terms of optical transition of electrons in typical degenerate semiconductors. The PL emission in the low energy region (0.8−1.2 eV) is attributed to the transition from the conduction band minima (CMB) (schematic in Fig. 5(c)).[2,33] The broad emission in the PL spectra (Fig. 5(a)) is attributed to the luminescence produced by the band filling effects (schematic in Fig. 5(c)).[33] The peak positions of broad luminescence are at 1.29 and 1.33 eV for samples A and D, respectively (Fig. 5(a)). Blue shift in the luminescence peaks for sample with increasing growth temperature is again attributed to band filling effect which is influenced by the carrier concentration present in the sample. As discussed earlier, the carrier concentration increases as the growth temperature increases. Consequently, electrons are prone to fill the high electronic levels in the conduction band. Moreover as a result of band filling, Fermi level is pinned further into the conduction band and it shifts close to the excitation energy. Thus, resonance conditions become prominent as the Fermi level approaches the excitation energy of 1.58 eV (785 nm). Consequently, resonance effects are prominent in high carrier concentration samples (C and D) because of increased closeness of filled levels towards the excitation energy. Thus, the forbidden mode of $B_1$(high), induced by near-resonance conditions, is more intense in high carrier concentration samples (C and D) than that for the sample A and B with low carrier concentration. However, carrier concentration dependent red shift of the forbidden mode frequency (Fig. 3) is yet to be understood.

**Conclusion**

InN nanoparticles are successfully prepared in chemical vapor deposition technique with the self-seeded catalytic approach. Increasing carrier concentration in the samples is achieved by increasing the growth temperature. Excitation dependent vibrational properties of InN nanoparticles are studied by Raman spectroscopy using different laser sources. With the excitation of 514.5 and 488 nm, Fano interference, instead of plasmon-phonon coupling, is analysed to dominate in high carrier concentration samples in case of $A_1$(LO) mode. Near-resonance excitation, invoking possible breakdown of the selection rule, induce significant changes in the Raman spectra with the appearance of the forbidden mode of $B_1$(high). Intensity and frequency of $B_1$(high) mode is observed to be strongly dependent on carrier concentration in the system. Thermal electrons also have its impression on the intensity of the $B_1$(high) mode along with inherent carrier concentration. Intensity of the forbidden mode is correlated with the band filling effects by studying the luminescence properties of InN nanoparticles.

**Acknowledgement**

We would like to acknowledge A. K. Tyagi, SND, IGCAR for his encouragement. We also thank Amirtha Pandian, and B. K. Panigrahi, Condensed Matter Physics Division, IGCAR for their help HRTEM studies, respectively.

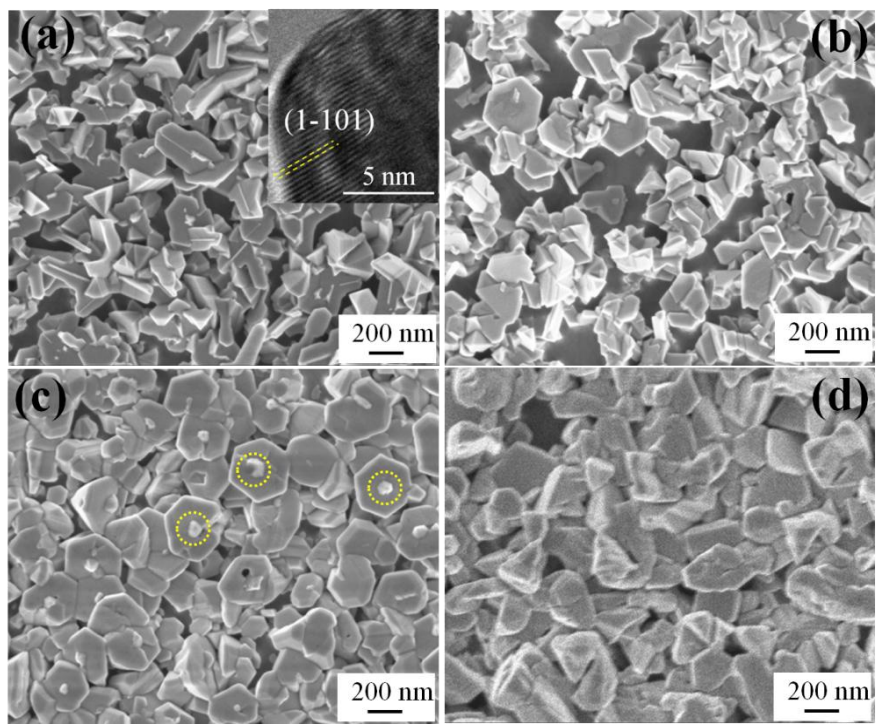

Fig. 1. FESEM micrographs of the InN nanostructures grown at different temperatures of (a) 580 °C showing random nanostructures. Inset shows the HRTEM image of one of the particle showing *m*-plane lattice fringes, (b) 600 °C showing similar random nanostructures, (c) 620 °C showing nanostructures with protruded hexagonal features (indicated by dashed circle) and (d) 650 °C showing random nanostructures.

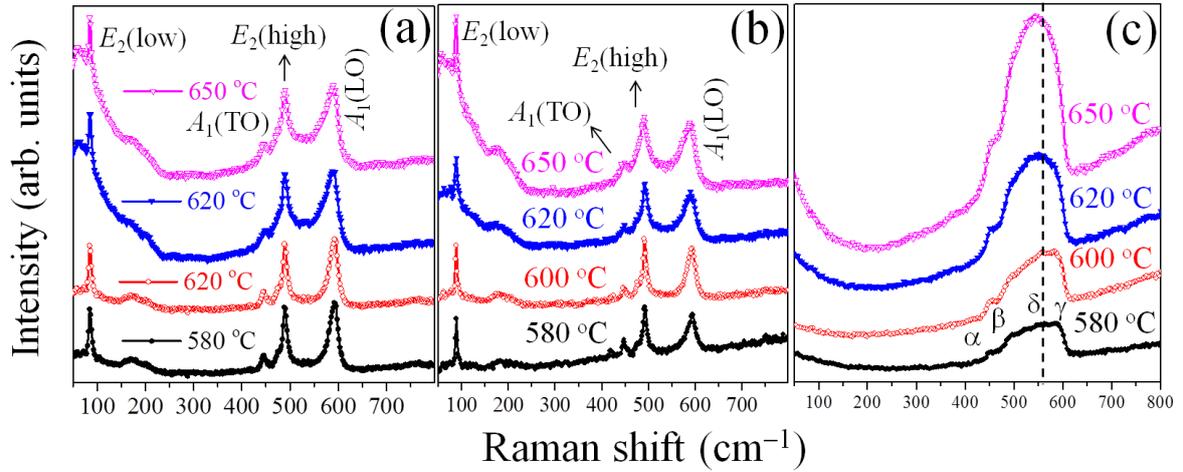

Fig. 2. Raman spectra collected for samples grown at different deposition temperatures with different excitations of (a) 514.5 nm, with high temperature (620 °C and 650 °C) grown samples showing the asymmetric broadening of $A_1$(LO) phonon mode because of Fano interference. (b) 488 nm, showing similar features as like 514.5 nm excitation (c) 785 nm, a broad feature dominating in the range of 440-590 cm$^{-1}$ with the signature of the symmetry allowed Raman modes at corresponding peak positions of α ($A_1$(TO)), β ($E_2$(high)), and γ ($A_1$(LO)) in addition to the forbidden mode, represented by δ ($B_1$(high)), dominating in high temperature grown samples. A dashed vertical line is indicated to show the red shift of the $B_1$(high) mode with increasing growth temperature.

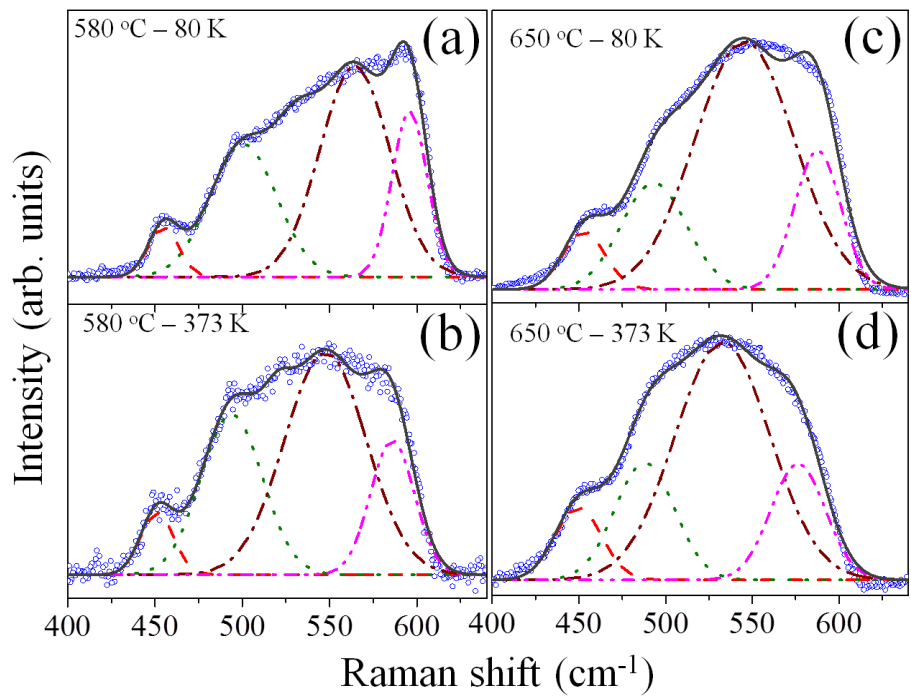

Fig. 3. Temperature dependent Raman spectra of InN nanoparticles grown at 580 and 650 °C with Gaussian fitting. Raman spectra collected at (a) 80K and (b) 373K for sample A with frequency of the forbidden mode at 565 and 544 cm$^{-1}$, respectively. Raman spectra collected at (c) 80K and (d) 373K for sample D with frequency of the forbidden mode at 552 and 536 cm$^{-1}$, respectively.

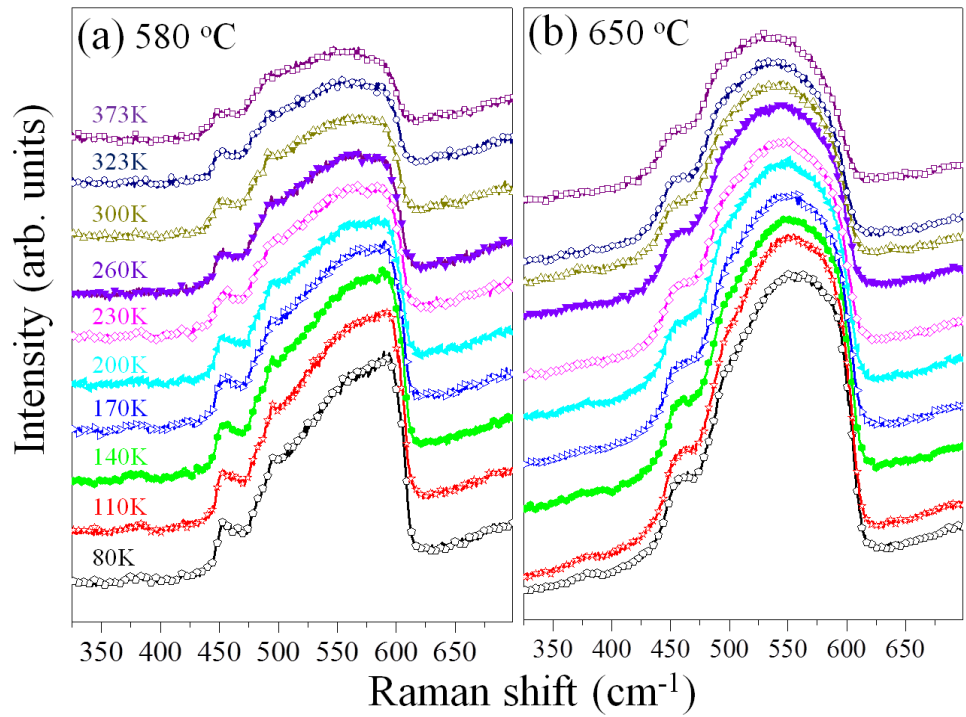

Fig. 4. Temperature dependent Raman spectra for (a) the sample A grown at 580 °C, with the forbidden mode dominating over $A_1$(LO) mode only at high temperatures and for (b) the sample D grown at 650 °C, with the forbidden mode dominating over $A_1$(LO) mode at all measurement temperatures.

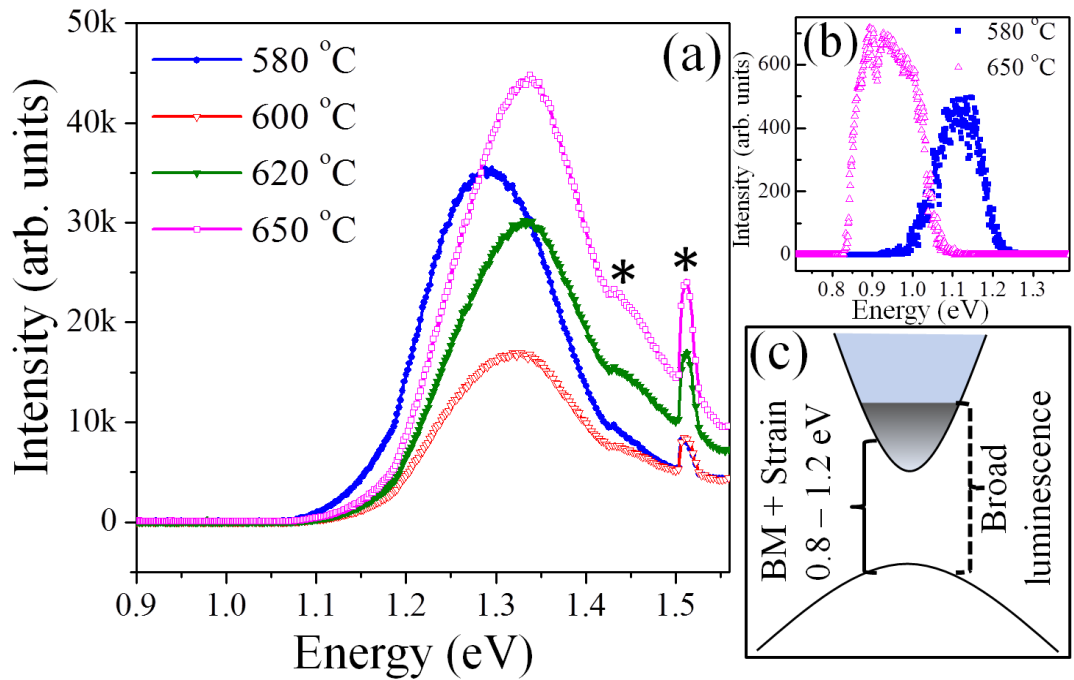

Fig. 5. (a) PL spectra collected for all samples showing the broad luminescence emission in the range of 1.2−1.4 eV along with the multi-phonon modes (indicated by *). A blue shift is observed in the broad emission peak with increasing temperature. (b) IR region of the PL spectra of InN grown at two extreme conditions, showing the near band edge emission in the range of 0.8 - 1.2 eV. (c) Schematic representation of band diagram depicting the band edge emission and the broad luminescence emission process.

**Supporting Information:**

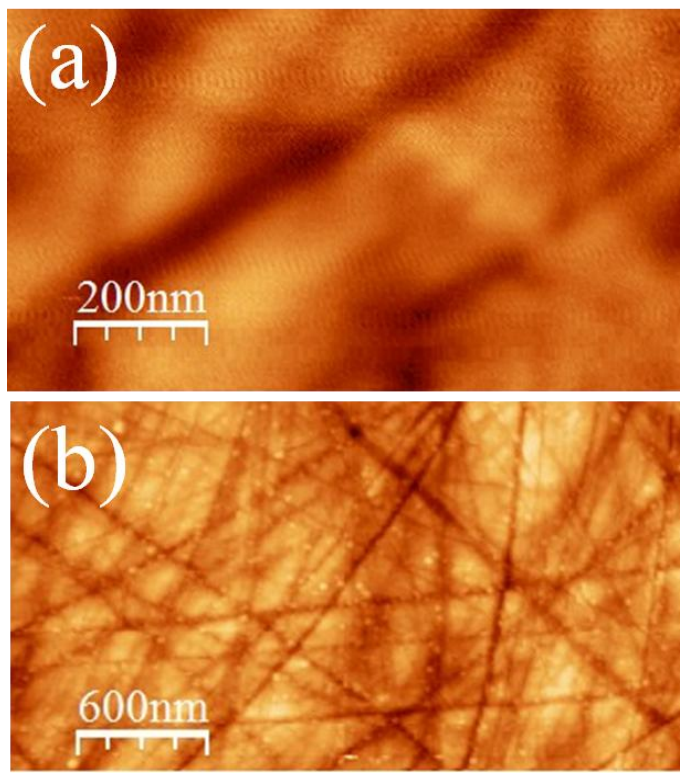

Figure S1. (a) Atomic force microscopic image of bare sapphire substrate before deposition. (b) In nanoparticles observed to be seeded after the temperature during growth raised to 500 °C and kept for 30 minutes.